\documentclass[12pt]{article}
\DeclareUnicodeCharacter{2212}{-}
\usepackage{url}
\usepackage{listings,jvlisting}
\usepackage{amssymb}
\usepackage{tabularx}
\usepackage{amsmath}
\usepackage{mdframed}
\usepackage{setspace}
\usepackage{threeparttable}
\usepackage[pdftex]{graphicx}
\usepackage{here}
\usepackage{authblk}
\usepackage{mathtools}
\usepackage{amsthm}
\makeatletter

\@addtoreset{equation}{section}
\makeatother

\makeatletter
\newcommand{\figcaption}[1]{\def\@captype{figure}\caption{#1}}
\newcommand{\tblcaption}[1]{\def\@captype{table}\caption{#1}}
\makeatother
\def\al{{\alpha}}

\def\de{{\delta}}

\def\th{{\theta}}

\def\bbe{{\text{\boldmath $\beta$}}}

\def\bep{{\text{\boldmath $\epsilon$}}}

\def\bth{{\text{\boldmath $\theta$}}}

\def\bmu{{\text{\boldmath $\mu$}}}

\def\bbeh{{\hat \bbe}}

\def\thh{{\hat \th}}

\def\bbeh{{\widehat \bbe}}

\def\bthh{{\widehat \bth}}

\def\bmuh{{\widehat \bmu}}

\def\bmut{{\widetilde \bmu}}

\def\btht{{\widetilde \bth}}

\def\Si{{\Sigma}}

\def\bSi{{\text{\boldmath $\Si$}}}

\def\e{{\text{\boldmath $e$}}}

\def\r{{\text{\boldmath $r$}}}

\def\v{{\text{\boldmath $v$}}}

\def\y{{\text{\boldmath $y$}}}

\def\G{{\text{\boldmath $G$}}}

\def\I{{\text{\boldmath $I$}}}

\def\K{{\text{\boldmath $K$}}}

\def\O{{\text{\boldmath $O$}}}

\def\R{{\text{\boldmath $R$}}}

\def\V{{\text{\boldmath $V$}}}

\def\X{{\text{\boldmath $X$}}}

\def\Z{{\text{\boldmath $Z$}}}
\def\thh{{\hat \th}}

\def\Bc{{\cal B}}

\def\Nc{{\cal N}}

\def\Uc{{\cal U}}

\def\[{{\text{\boldmath $[$}}}
\def\]{{\text{\boldmath $]$}}}

\def\zero{{\bf\text{\boldmath $0$}}}

\def\|{{\,|\,}}

\def\/{{\Bigr/\!\!}}

\def\1r{{\rm (1)}}
\def\2r{{\rm (2)}}
\def\3r{{\rm (3)}}
\def\4r{{\rm (4)}}
\def\5r{{\rm (5)}}

\def\non{{\nonumber}}

\begin{document}
\title{Easy confidence interval formulas for network meta-analysis and adjustment of confidence intervals for a small number of studies}
\author[1,2]{Masahiro Kojima\footnote{Address: Biometrics Department, R\&D Division, Kyowa Kirin Co., Ltd.
Otemachi Financial City Grand Cube, 1-9-2 Otemachi, Chiyoda-ku, Tokyo, 100-004, Japan. Tel: +81-3-5205-7200 \quad
E-Mail: masahiro.kojima.tk@kyowakirin.com}}
\affil[1]{Kyowa Kirin Co., Ltd}
\affil[2]{The Graduate University for Advanced Studies}

\maketitle

\abstract{\noindent
We propose simple formulas of confidence intervals for the Wald statistic, likelihood ratio statistic, and score statistic for a network meta-analysis. In addition, we consider resolutions for concerns that network meta-analyses with a small number of studies cannot hold a nominal confidence level. For a bias adjustment in analyses with a small number of studies, a Bartlett-type adjustment is a well-known method. Many Bartlett-type adjustment-type methods are based on maximum likelihood estimators. However, the network meta-analysis often uses the restricted maximum likelihood estimators that have not been extensively discussed in Bartlett-type adjustment.\\
In this paper, we propose a Bartlett-type adjustment method for the Wald statistic, likelihood ratio statistic, and score statistic when nuisance parameters are estimated by not only the maximum likelihood method but also the restricted maximum likelihood method. In addition, we propose a higher-order adjustment by applying the bootstrap method to the Bartlett-type adjusted statistics. Using a computer simulation, we confirmed that the adjusted confidence intervals maintained a nominal confidence level. In addition, we confirmed that the confidence interval of the likelihood ratio test based on the restricted maximum likelihood method performs well without further bootstrap adjustment. Finally, we demonstrated that confidence intervals were adjusted for actual network meta-analysis.
}
\par\vspace{4mm}
\section{Introduction}
\label{s:intro}

Over the past few decades, many researchers have shown an interest in network meta-analysis (NMA) which has been actively used in a wide range of fields including ecology, economics, medical science, and medicine. As Gurevitch et al.\cite{Gurevitch2018} indicated, studies using the meta-analysis began in the 1970s; in the recent years, the number of studies has rapidly increased and currently exceeds 200,000 papers. The number of papers using network meta-analyses has also grown\cite{Li2016}. With the number of clinical trials increasing each year\cite{Viergever2015}, studies using the network meta-analysis seem likely to be more numerous in the future.

NMA synthesizes treatment effects obtained from high-quality studies via systematic reviews\cite{Jackson2011}. There may be discrepancies in the results of the studies to be synthesized, and such discrepancies are called heterogeneity. To address the problem of the heterogeneity, a random effect is used to adjust heterogeneities between studies. In the NMA, the random effects meta-analysis model\cite{White2012} is often used\cite{Tricco2017,Brunoni2017,Mitra2018}.

Here, we consider the NMA with a small number of studies. The problem in such cases is that the confidence intervals cannot hold a nominal confidence level ($100-\al$)$\%$. Consequently, the confidence intervals become wider or narrower than the true confidence intervals, potentially resulting in misleading decision-making. To resolve this problem with small numbers of studies, the Bartlett-type adjustment method is often used. Because the Bartlett-type adjustment can adjust the second-order bias in type I error\cite{Rothenberg1984,Cordeiro2014}, the accuracy of confidence intervals is improved.  Many Bartlett-type adjustment methods are based on maximum likelihood estimators (MLE). However, the NMA often uses the restricted maximum likelihood estimators (REMLE), which have not been extensively discussed in this context. Noma et al.\cite{Noma2018} proposed the Bartlett-type adjustment using $m$ bootstrap samples for the NMA with a small number of studies. However, there is a problem with the bootstrap samples, i.e., the calculation load is very high because the MLE or REMLE is derived by numerically solving a likelihood equation that is not closed-form. In addition, because the likelihood ratio (LR) statistic and score statistic on the REMLE are not closed-form, those confidence intervals also need to be derived numerically. When using the bootstrap method, we do not know which bias terms cause the bias. Kojima and Kubokawa\cite{Kojima2013} analytically derived Bartlett-type adjustment terms in general linear models via an analytical approach. However, they did not propose a Bartlett-type adjustment method based on the restricted likelihood function. van Aert and Jackson\cite{Aert2019} proposed using t-quantiles instead of normal quantiles to improve the actual coverage probability to be closer to the nominal level. However, their proposed method cannot be applied the NMA straightforwardly.

In this paper, we propose novel Bartlett-type adjustment terms for confidence intervals based on the Wald, LR, and score using the MLE and REMLE. Because the Bartlett adjustment terms can be calculated analytically, our proposed method can calculate the adjusted confidence intervals at 1/$m$ of the speed of the bootstrap method. In addition, we propose the closed-form confidence intervals of LR and score statistic based on a novel constrained likelihood function under a null hypothesis. Thus, our method does not need to use numerical analysis except for the derivation of the MLE. We focus on the likelihood-based inference methods to derive the equations for adjustment.

We give further details of the Bartlett adjustment and Bartlett-{\bf type} adjustment. Here, "Bartlett adjustment" refers to the adjustment only for the likelihood ratio statistics. In general, the Bartlett adjustment terms can be derived from the expectation of the LR statistic. By contrast, the "Bartlett-{\bf type} adjustment" is an extension of the Bartlett adjustment that also supports the adjustment of the Wald statistic and score statistic. In other words, the Bartlett-type adjustment method is more generalized method. The Bartlett-type adjustment terms can be derived from the asymptotic expansion of the statistic.

In Section 2, we introduce the random effects meta-analysis model and the three statistics. In addition, we provide the formulas for confidence intervals. In Section 3, we obtain the Bartlett-type adjustment terms via the analytical approach and the adjusted confidence intervals. In Section 4, we propose the higher-order adjustment using the parametric bootstrap method. In Section 5, we use a computer simulation to confirm that the adjusted confidence intervals maintained a nominal confidence level. In Section 6, we demonstrate that confidence intervals were adjusted for actual network meta-analysis. Section 7 is the discussion. All technical issues are given in the Appendix in the supplemental material.

\section{Methods}
\section{Three statistics in the random effects meta-analysis model}\label{sec2}
\subsection{Random effects meta-analysis model}\label{subsec21}

We assume that the number of studies is $n$ and the number of treatment effects of interest is $p$. The treatment effects $p$ are differences of outcomes between treatment groups (e.g., any pairwise comparison) and a reference treatment (e.g., a placebo). The effect measures are a mean difference, log odds ratio difference, and log hazard ratio\cite{Bruce2003,Stettler2007,Eliott2007,Lee2017}. In the NMA, because the every study may not include the all treatment groups, we define that the number of outcomes of the $i$-th study is $p_i$ independent contrast estimates and $p_{i}\leq p$ satisfies. We assume $p$ is a finite, $N=\sum^{n}_{i=1}p_{i}$, and $N>p$. 

We consider the random effects meta-analysis model consisting of the within-study model and between-studies model\cite{Jackson2011,Noma2018}, 
\begin{align}
\mbox{(within-study model) }\y_{i}=&\X_{i}\bbe_{i}+\e_ {i},\non\\
\mbox{(between-studies model) }\bbe_{i}=&\bmu+\v_i,\non
\end{align}
where $\y_{i}$ is a $p_{i}\times 1$ vector of a known estimator of a treatment effect within the $i$-th study, $\X_{i}$ is a known $p_{i}\times p$ design matrix in which the element of a treatment corresponding to $y_{ij}$ is 1 and the others are 0, $\bbe_{i}$ is a $p\times 1$ vector of a treatment effect within the $i$-th study, $\e_{i}$ is a $p_{i}\times 1$ vector of an error within the $i$-th study$,\e_{i}\sim\Nc(\zero, \R_{i})$, $\R_{i}$ is a known $p_{i}\times p_{i}$ covariance matrix of the $i$-th study, $\bmu$ is a $p\times 1$ parameter vector of the average treatment effect across all studies, $\v_i$ is a $p\times 1$ vector of an error between studies, $\v_i\sim\Nc(\zero,\V(\bth))$, $\V(\bth)$ is a $p\times p$ covariance matrix, and $\bth$ is a $q\times 1$ nuisance unknown parameter vector. The covariance structure of $\V(\bth)$ is known. The treatment effects $\bbe_i$ within the $i$-th study are extracted $p_i$ parameters by the design matrix $\X_i$. 

To easily calculate the Bartlett-type adjustment, we consider the marginal model of the random effects meta-analysis model as
\begin{equation}
\y=\X\bmu+\bep,\hspace{10pt}\bep\sim\Nc(\zero,\bSi(\bth)),\hspace{10pt}\bSi(\bth)=\R+\G(\bth) ,
\label{eqn:model}
\end{equation}
where $\y=(\y_{1}^T,\ldots,\y_{n}^T)^T$ is an $N\times 1$ vector, $\X=(\X_{1}^T,\ldots,\X_{n}^T)^T$ is an $N\times p$ matrix, $\R$=diag$(\R_1,\ldots,\R_n)$ is a $N\times N$ matrix, $\G(\bth)=\Z$diag$(\V(\bth),\ldots,\V(\bth))\Z^T$ is an $N\times N$ matrix,  $\Z$=diag$(\X_1,\ldots,\X_n)$ is an $N\times np$ matrix. diag denotes the block diagonal matrix. We assume rank$(\X)=p$.

We provide a simple example of the model. There are three studies in the NMA. The first study has three treatments (drug A, drug B, and placebo). The second study has two treatments (drug A and placebo). The third study has two treatments (drug B and placebo). The difference of A and P is $y_{A-P}$, the difference of B and P is $y_{B-P}$, the standard error of the difference of A and P is $s_{A-P}$, the standard error of the difference of A and P is $s_{B-P}$. For simplicity, this example does not consider the correlation of $A-P$ and $B-P$ and the difference between A and B. In the first study, $\y_1=(y_{A-P},y_{B-P})^T$, $\X_1=((1,0),(0,1))$, $\R_1=((s^2_{A-P},s_{A-P}s_{B-P}/2),(s_{A-P}s_{B-P}/2,s^2_{B-P}))$, $\V(\bth)=((\th_1,\th_3),(\th_3,\th_2))$. In the second study, $y_2=y_{A-P}$, $\X_2=(1,0)$, $\R_2=s^2_{A-P}$. In the third study, $y_3=y_{B-P}$, $\X_3=(0,1)$, $\R_3=s^2_{B-P}$. The marginal model of the random effects meta-analysis model is obtained by combining the information of each study.

In general, the NMA makes evaluations using the confidence intervals of $\mu_{i}$. The confidence intervals based on LR statistic and Score statistic need the null hypothesis\cite{Noma2018,Sugasawa2021}. The null hypothesis is defined, 
\begin{equation}
H_0\ :\ \r^T\bmu=\r^T\bmu_0,
\label{eqn:hyp}
\end{equation}
where $\r$ and $\bmu_0$ are known $p\times 1$ constant vectors, the $i$-th element of $\r$ is $1$ and the others are $0$. The element $1$ of $\r$ means to indicate the confidence interval for the parameter of interest. If we are interested in differences between treatments, we set $\bmu=\zero$. If you are interested in whether the confidence interval is included for some clinically meaningful threshold, put the clinically meaningful threshold in $\bmu_0$. If we are interested in whether or not a confidence interval includes for some clinically meaningful value, we put a clinically meaningful threshold in $\bmu_0$.

\subsection{Wald, LR, and Score statistic}\label{subsec22}
We introduce the Wald, LR, and Score statistics. The three statistics consist of two types of the log likelihood function, the general log likelihood function and the constrained log likelihood function under the null hypothesis. 

First of all, the log likelihood function is
\begin{equation}
l^{L}(\bmuh(\bth),\bth)=C_1-\frac{1}{2}\mbox{ln}|\Si(\bth)|-\frac{1}{2}\left(\y-\X\bmuh(\bth)\right)^T\Si^{-1}(\bth)\left(\y-\X\bmuh(\bth)\right),
\label{eqn:MLF}
\end{equation}
where $C_1$ is a constant value and $\bmuh(\bth)=\left(\X'\Si^{-1}(\bth)\X\right)^{-1}\X'\Si^{-1}(\bth)\y$ is the maximum likelihood estimator derived by maximizing $l^{L}(\bmu,\bth)$ about $\bmu$. The MLE of $\bth$ is derived by maximizing (\ref{eqn:MLF}). Next, we consider the constrained log likelihood function under the null hypothesis, which can be derived using the method of the Lagrange multipliers about $l^{L}(\bmu,\bth)$. The constrained log likelihood function under the null hypothesis has $l^{L}(\bmut(\bth),\bth)$, where
\begin{equation}
\bmut(\bth)=\bmuh(\bth)-\frac{\left(\r^T\bmuh(\bth)-\r^T\bmu_0\right)}{\r^T(\X^T\Si^{-1}(\bth)\X)^{-1}\r}(\X^T\Si^{-1}(\bth)\X)^{-1}\r.
\label{eqn:RMLM}
\end{equation}
The constrained log likelihood function under the null hypothesis can be expressed by replacing $\bmuh(\bth)$ of (\ref{eqn:MLF}) with $\bmut(\bth)$, and this relationship is important to derive the Bartlett-type adjustment terms. The MLE of $\bth$ under the null hypothesis is derived by maximizing $l^{L}(\bmut(\bth),\bth)$. 

In the NMA, the REMLE is often used. We assume that $\K$ is a function of the $\X$ satisfying $\K\X=\O$ and rank($\K$)= $N-$rank$(\X)$. The restricted likelihood of data $y$ is defined as the likelihood of data $\K\y$, the restricted log likelihood function is shown as $l^{RL}(\bmuh(\bth),\bth)=C_2-\frac{1}{2}\mbox{ln}|X^T\Si^{-1}(\bth)X|+l^{L}(\bmuh(\bth),\bth)$\cite{Welham1997,Jennrich1986}, 
where $C_2$ is a constant value. The REMLE of $\bth$ is derived by maximizing $l^{RL}(\bmuh(\bth),\bth)$. We consider the constrained log likelihood function under the null hypothesis. Since the restricted log likelihood function does not include $\bmu$, we cannot define the constrained log likelihood function under the null hypothesis in the same way as the log likelihood function $l^{L}(\bmut(\bth),\bth)$. Welham and Thompson\cite{Welham1997} show the two types of constrained log likelihood function under the null hypothesis. The first is the sub-model and The second is dropped terms relating to the null hypothesis. However, the both restricted log likelihood functions cannot lead to the closed-form confidence intervals for the LR and Score statistics. In this paper, we refer to Welham and Thompson\cite{Welham1997}'s function dropping terms, which assumes a part of $\bmu$ being $\zero$. We define a new extension of Welham and Thompson\cite{Welham1997}'s function for leading the closed-form confidence intervals. Now, we substitute $\bmu$ into $\bmuh(\bth)$ of $l^{RL}(\bmuh(\bth),\bth)$ and we obtain the constrained log likelihood function under the null hypothesis in the same manner as the log likelihood function. Then, the new constrained log likelihood function under the null hypothesis can be expressed as $l^{RL}(\bmut(\bth),\bth)$, and the REMLE of $\bth$ under the null hypothesis can be obtained. The REMLE of $\bth$ is confirmed to have consistency in Appendix A.10. 

From here, when it is not necessary to distinguish $l^{L}$ from $l^{RL}$, we write just $l$. In addition, we define that $\bthh$ is the MLE or REMLE, and that $\btht$ is  the MLE or REMLE under the null hypothesis.

We introduce the three statistics using the difference between two log likelihood functions.

{\bf [1] Wald statistic}\ \
\begin{align}
W(\bthh)=-2\left(l(\bmut(\bthh),\bthh)-l(\bmuh(\bthh),\bthh)\right)=\frac{\left(\r^T\bmuh(\bthh)-\r^T\bmu_0\right)^2}{h(\bthh)^2}, \non
\end{align}
where $h(\bth)=\sqrt{\r^T(\X^T\bSi(\bth)^{-1}\X)^{-1}\r}$. The derivation of the last term is shown in Appendix A.3. From the last term, $W(\bthh)$ is the closed-form. Thus, we can easily derive the confidence interval.

{\bf [2] LR statistic}\ \
\begin{align}
LR(\bthh,\btht)=&-2\left(l(\bmut(\btht),\btht)-l(\bmuh(\bthh),\bthh)\right)=-2\de(\bthh,\btht)+W(\btht), \non
\end{align}
because $l(\bmut(\btht),\btht)-l(\bmuh(\bthh),\bthh)=\de(\bthh,\btht)+\left(l(\bmut(\btht),\btht)-l(\bmuh(\btht),\btht)\right)$ and Appendix A.1. $\de(\bthh,\btht)=l(\bmuh(\btht),\btht)-l(\bmuh(\bthh),\bthh)=O_p(N^{-1})$ since it can be proved in Appendix A.7 and \cite{Kubokawa2011}. The $\bmu_0$ is only included in $W(\btht)$, $W(\btht)$ is the closed-form about $\bmu_0$. Thus, we can derive the closed-form confidence interval of $LR(\bthh,\btht)$. 
Hardy and Thompson\cite{Hardy1996} propose a confidence interval of likelihood ratio statistic using a likelihood function. However, The confidence interval is calculated to satisfy $l(\bmu,\bthh)>l(\bmuh(\bthh),\bthh)-3.84/2$, which increases the computational load. We consider the simple confidence interval of likelihood ratio statistic based on the null hypothesis $\r^T\bmu=\r^T\bmu_0$.

{\bf [3] Score statistic}\ \
\begin{equation}
S(\btht)=-2\left(l(\bmut(\btht),\btht)-l(\bmuh(\btht),\btht)\right)=\frac{(\r^T\bmuh(\btht)-\r^T\bmu_0)^2}{h(\btht)^2}=W(\btht). \non
\end{equation}
The Score statistic can be defined by replacing $\bthh$ of the Wald statistic with $\btht$. Therefore, because $S(\btht)$ can be expressed as closed-formed $W(\cdot)$, we can formulate the three confidence intervals.

We formulate the confidence intervals. These three statistics are asymptotically distributed as $\chi^2_1$. The coverage probability of confidence interval of the Wald statistic is shown as
\begin{align}
P\left(W(\bth)\leq z_{\al\%,\chi^2}\right)&= P\left(-\sqrt{z_{\al\%,\chi^2}}\leq\frac{\r^T\bmuh(\bthh)-\r^T\bmu}{h(\bthh)}\leq \sqrt{z_{\al,\chi^2}}\right)\non\\
&=P\left(\r^T\bmuh(\bthh)-z_{\al\%/2}h(\bthh)\leq\r^T\bmu\leq \r^T\bmuh(\bthh)+z_{\al/2}h(\bthh)\right),\non
\end{align}
where $z_{\al\%, \chi^2}$ is the upper $\al\%$ of $\chi^2_1$, $z_{\al\%/2}$ is the upper $\al\%/2$ of the standard normal distribution and $z_{\al\%, \chi^2}=z_{\al\%/2}^2$. Since the confidence intervals are not based on the null hypothesis, $\bmu_0$ in each statistic must be replaced with $\bmu$. Then, based on $W(\cdot)$, the confidence intervals of the other statistics are also given below.

\medskip
{\bf Confidence interval based on the Wald statistic:}\ \
\begin{align}
\r^T\bmuh(\bthh)-z_{\al/2}h(\bthh)\leq\r^T\bmu\leq\r^T\bmuh(\bthh)+z_{\al/2}h(\bthh).
\label{eqn:WCI}
\end{align}

\medskip
{\bf Confidence interval based on the LR statistic:}\ \
\begin{align}
\r^T\bmuh(\btht)-\sqrt{z_{\al/2}^2+2\de(\bthh,\btht)}h(\btht)\leq\r^T\bmu\leq\r^T\bmuh(\btht)+\sqrt{z_{\al/2}^2+2\de(\bthh,\btht)}h(\btht).
\label{eqn:LRCI}
\end{align}

\medskip
{\bf Confidence interval based on the Score statistic:}\ \ 
\begin{align}
\r^T\bmuh(\btht)-z_{\al/2}h(\btht)\leq\r^T\bmu\leq\r^T\bmuh(\btht)+z_{\al/2}h(\btht).
\label{eqn:SCI}
\end{align}

We show numerical examples in the supplementary materials.

Noma et al.\cite{Noma2018} do not show the closed-form confidence interval of the LR and Score statistics. However, we can derive the closed-form confidence intervals of all statistics based on $W(\cdot)$. On the other hand, the three confidence intervals cannot hold the nominal confidence level when the number of studies is small because the approximation accuracy of the statistics to $\chi^2_1$ depends on the number of studies. To be strict, distribution functions of the statistics have a bias term of the second order. In other words, in the case of Wald, the distribution function is $P\left(W(\bthh)\leq z_{\al,\chi^2}\right)=1-\al+O\left(N^{-1}\right)$, which has the bias term($O\left(N^{-1}\right)$). In the next chapter, we derive the asymptotic expanded distribution function and the Bartlett adjustment-type terms of the second-order bias.

\section{Bartlett-type adjustment}\label{sec3}

The Bartlett-type adjustment is a method that removes the bias of the second-order. If we apply the Bartlett-type adjustment to the statistic, $P\left(W_{ad}(\bthh)\leq z_{\al,\chi^2}\right)=1-\al+O\left(N^{-3/2}\right)$ where $W_{ad}(\bthh)$ is the adjusted statistic. We show the technical notation and assumption of $\bthh$ in the supplemental material.

Firstly, we consider to expanding asymptotically the distribution function of the Wald statistic. Kojima and Kubokawa\cite{Kojima2013} have already derived the asymptotic expansion of the distribution function of the Wald statistic. However, it was expressed as a matrix form and is complicated to understand. In this study, we re-calculate it in a vector form to simplify some equations and make it easy to understand. 

The original Wald statistic $W(\bthh)$ is difficult to asymptotic, thus we divide it into three parts, 
\begin{equation}
W(\bthh)=\frac{(a+b)^2}{1-c},\non
\end{equation}
where
\begin{align}
&a=\frac{\r^T\bmuh(\bthh)-\r^T\bmuh(\bth)}{h(\bth)},\hspace{2pt}b=\frac{\r^T\bmuh(\bth)-\r^T\bmu_0}{h(\bth)},\hspace{2pt}c=-\frac{h(\bthh)^2-h(\bth)^2}{h(\bth)^2}.\non
\end{align}
From the assumption and Taylor series, we have $a=a_1+O_p(N^{-1})$, where $a_1=O_p(N^{-1/2})$, and $c=c_1+c_2+O_p(N^{-3/2})$, where $c_1=O_p(N^{-1/2})$, $c_2=O_p(N^{-1})$. We show the technical description to derive the Bartlett-type adjustment in the supplementary materials. The Bartlett-type adjustment $w_{ad}(x)$ satisfies $P\left(W(\bthh)\leq w_{ad}(x)x\right)=F_{\chi^2_1}(x)+O(N^{-3/2})$ where
\begin{align}
w_{ad}(x)=1+\widehat{E[a_1^2]}+\widehat{E[c_2]}+\frac{1}{4}(1+x)\widehat{E[c_1^2]}.
\label{eqn:Wcor}
\end{align}
The $\widehat{E[a_1^2]}$, $\widehat{E[c_2]}$, and $\widehat{E[c_1^2]}$ are plugged $\bthh$ to $E[a_1^2]$, $E[c_2]$, and $E[c_1^2]$, respectively. $\widehat{E[a_1^2]}$ means the bias of $\bbeh(\bthh)$. $\widehat{E[c_2]}$ and $\widehat{E[c_1^2]}$ mean the biases of $h(\bthh)$. To understand the bias terms, we consider the simplest case which is $\bSi(\bth)=\th_1 \I_N$. In this case, $\widehat{E[a_1^2]}$ is zero because $\bmuh(\bth)$ does not include $\bth$, $\widehat{E[c_2]}$ is zero because $h(\bth)^2$ is a linear function about $\bth$, and $\widehat{E[c_1^2]}$ is $\frac{1}{2N}$. Therefore, from (\ref{eqn:Wcor}), the Wald statistic has a bias of $\frac{1}{8N}(1+x)$. However, in this study, $\bSi(\bth)$ becomes more complicated, and each bias term may be larger depending on $\bth$.

Next, we derive the Bartlett-type adjustment terms of the LR statistic. By using the Taylor expansion of $l(\bmut(\bthh),\bthh)$ of the LR statistic around $\btht$, the expanded LR statistic is obtained (details in Appendix A.9). The characteristic function of the LR statistic is given using Appendix A.10. The expanded distribution function of the LR statistic is
\begin{align}
P(LR(\bthh,\btht)\leq x)=&F_{\chi^2_1}(x)-x\left(E[c_2]+\frac{1}{4}E[c_1^2]\right)f_{\chi^2_{1}}(x)+O(N^{-3/2}).\non
\end{align}
There is no bias term of $\bmuh(\bthh)$ in the second-order. In addition, the weight of the bias $E[c_1^2]$ is reduced to $\frac{x}{4}$ compared with the Wald statistic. Therefore, the LR statistic is expected to be less biased than the Wald statistic. Besides, because $\bthh^{REML}$ has been shown to be less the bias asymptotically than $\bthh^{ML}$ by Kubokawa\cite{Kubokawa2011}, the LR statistic with the REMLE is considered to be less biased.

We obtain the Bartlett-type adjustment terms, 
\begin{align}
lr_{ad}(x)=1+\widehat{E[c_2]}+\frac{1}{4}\widehat{E[c_1^2]}.
\label{eqn:LRcor}
\end{align}

Finally, we derive the Bartlett-type adjustment terms of the Score statistic. The Score statistic is asymptotically expressed by the Taylor expansion $l(\bmut(\bthh),\bthh)$ of the LR statistic around $\btht$, yielding
$S(\btht)=W(\btht)=W(\bthh)-\frac{1}{2}\sum_{i,j=1}^q\thh_{1i}\thh_{1j}(2a_{1i}b-b^2c_{1i})(2a_{1j}b-b^2c_{1j})+O_p(N^{-3/2})$.
The characteristic function is obtained using Appendix A.10. Thus, calculating the distribution function using the inversion formula, we obtain the Bartlett-type adjustment terms where
\begin{align}
s_{ad}(x)=1-\widehat{E[a^2_1]}+\widehat{E[c_2]}+\frac{1}{4}(1-x)\widehat{E[c_1^2]}.
\label{eqn:Scor}
\end{align}
The Score statistic also includes the bias $\widehat{E[a_1^2]}$ of $\bmuh(\bthh)$, and the extra bias of the variance is similar to that of the Wald statistic. We consider why Wald and Score statistics include the extra biases relative to the LR statistic. The Wald and Score statistics have the mismatched estimator $\bmut(\bthh)$ or $\bmuh(\btht)$, which have mixed estimators calculated under the non-null and null hypotheses. Therefore, the extra biases occur. On the other hand, the LR statistic is that there is no problem in the combination of estimators, and the bias is small. We show the formulas of each adjustment terms in Appendix 14.

\begin{mdframed}
\noindent 
\textbf{\textit{Theorem} 1.}
\\
Following the assumption in appendix A.2, the Bartlett-type adjusted three statistics are (\ref{eqn:Wcor}), (\ref{eqn:LRcor}), and (\ref{eqn:Scor}), respectively. Then, we obtain the adjusted closed-form confidence intervals below.

\medskip
\noindent
{\bf Confidence interval of the Bartlett-type adjusted Wald statistic.}\ \
\begin{align}
\r^T\bmuh(\bthh)-z_{\al/2}h(\bthh)\sqrt{w_{ad}(z_{\al/2}^2)}\leq\r^T\bmu\leq\r^T\bmuh(\bthh)+z_{\al\%/2}h(\bthh)\sqrt{w_{ad}(z_{\al/2}^2)}
\label{eqn:WcCI}
\end{align}

\medskip
\noindent
{\bf Confidence interval of the Bartlett-type adjusted LR statistic.}\ \
\begin{align}
\r^T\bmuh(\btht)-\sqrt{LR_{ad}}h(\btht)\leq\r^T\bmu\leq\r^T\bmuh(\btht)+\sqrt{LR_{ad}}h(\btht),
\label{eqn:LRcCI}
\end{align}
where $LR_{ad}=z_{\al/2}^2lr_{ad}(z_{\al/2}^2)+2\de(\bthh,\btht)$.

\medskip
\noindent
{\bf Confidence interval of the Bartlett-type adjusted Score statistic.}\ \ 
\begin{align}
\r^T\bmuh(\btht)-z_{\al/2}h(\btht)\sqrt{s_{ad}(z_{\al/2}^2)}\leq\r^T\bmu\leq\r^T\bmuh(\btht)+z_{\al/2}h(\btht)\sqrt{s_{ad}(z_{\al/2}^2)}
\label{eqn:ScCI}
\end{align}
The coverage probabilities in the above adjusted confidence intervals are $1-\al+O(N^{-3/2})$, thus the second-order bias is removed.
\end{mdframed}

We show an example of the Bartlett-type adjustment terms by using the assumptions used in the the confidence interval example in the section \ref{sec2}. $\widehat{E[a^2_1]}=0$, $\widehat{E[c_1^2]}=\frac{n}{(1+\thh)^4}$, $\widehat{E[c_2]}=\frac{n^2}{2(1+\thh)^4}$ with the MLE, and $\widehat{E[c_2]}=0$ with the REMLE.

\section{Higher order adjustment via parametric bootstrap}\label{sec4}
We consider whether further adjustments can be made to the adjusted statistics. Kojima and Kubokawa\cite{Kojima2013} and Noma et al.\cite{Noma2018} propose a bias adjustment method using a parametric bootstrap. The parametric bootstrap generates the bootstrap samples of $m$ sets from a parametric distribution. The statistics using each bootstrap sample are calculated, we obtain the upper $\al\%$ of the sorted $m$ statistics as the rejection point. By using this rejection point, the accuracy is impoved.

In this paper, we consider to use the parametric bootstrap method to $P\left(W_{ad}(\bthh)\leq z_{\al\%,\chi^2}\right)=1-\al+O\left(N^{-3/2}\right)$, thus we have $P\left(W_{ad}(\bthh)\leq z_{\al\%,w}^{*}\right)=1-\al+o\left(N^{-3/2}\right)$ where $*$ refers to bootstrap samples and $z_{\al\%,w}^{*}$ is the upper $\al\%$ of $W^{*}_{ad}(\bthh^{*})$.  We propose the procedure for the bootstrap adjustment method below.

\medskip
\noindent
[Step 1.]\\
Generate the parametric bootstrap samples $\y^{*}$ of $m$ sets from $\X\bmut(\bthh)+\bep^{*}$ where \cite{Kojima2013} and \cite{Noma2018} propose to generate using $\bmut$ constrained under the null hypothesis.

\medskip
\noindent
[Step 2.]\\
Calculate the $m$ Bartlett-type adjusted statistics.

\medskip
\noindent
[Step 3.]\\
Select the upper $\al\%$ of {\it m} of sorted statistics, and find the square root.

\begin{mdframed}
\textbf{\textit{Theorem} 2.}
\\
Following the assumption in appendix A.2, the Bartlett-type adjusted three statistics are (\ref{eqn:Wcor}), (\ref{eqn:LRcor}), and (\ref{eqn:Scor}), respectively. The upper $\al\%$ of the three statistics via the parametric bootstrap method are $z_{\al\%,w}^{*}$, $z_{\al,LR}^{*}$, and $z_{\al,S}^{*}$. Then, we obtain the adjusted closed-dorm confidence interval using the bootstrap adjustment method,

\medskip
\noindent
{\bf Bootstrap Confidence interval of the Bartlett-type adjusted Wald statistic}\ \
\begin{align}
\r^T\bmuh(\bthh)-h(\bthh)\sqrt{z_{\al\%,w}^{*}w_{ad}(z_{\al/2}^2)}\leq\r^T\bmu\leq\r^T\bmuh(\bthh)+z_{\al/2}h(\bthh)\sqrt{z_{\al\%,w}^{*}w_{ad}(z_{\al/2}^2)}\non
\end{align}

\medskip
\noindent
{\bf Bootstrap Confidence interval of the Bartlett-type adjusted LR statisticR}\ \
\begin{align}
\r^T\bmuh(\btht)-\sqrt{LR^{\ast}_{ad}}h(\btht)\leq\r^T\bmu\leq\r^T\bmuh(\btht)+\sqrt{LR^{\ast}_{ad}}h(\btht),\non
\end{align}
where $LR^{\ast}_{ad}=z_{\al,lr}^{*}lr_{ad}(z_{\al/2}^2)+2\de(\bthh,\btht)$.

\medskip
\noindent
{\bf Bootstrap Confidence interval of the Bartlett-type adjusted Score statistic}\ \ 
\begin{align}
\r^T\bmuh(\btht)-h(\btht)\sqrt{z_{\al,s}^{*}s_{ad}(z_{\al/2}^2)}\leq\r^T\bmu\leq\r^T\bmuh(\btht)+h(\btht)\sqrt{z_{\al,s}^{*}s_{ad}(z_{\al/2}^2)}\non
\end{align}
The coverage probabilities of (\ref{eqn:WcCI}), (\ref{eqn:LRcCI}), and (\ref{eqn:ScCI}) are $1-\al+O(N^{-3/2})$ The coverage probabilities in the bootstrap confidence intervals are $1-\al+o(N^{-3/2})$, thus the third-order bias is removed.
\end{mdframed}

\section{Simulation study}\label{sec6}
We perform a computer simulation to confirm the accuracy of the Bartlett-type adjusted statistics through two simulation studies. We generate the datasets using the same procedure as Sidik and Jonkman\cite{Sidik2007}, Noma et al. \cite{Noma2018}, and Nagashima et al. \cite{Nagashima2019}.
\subsection{Simulation 1}\label{subsec61}
\begin{figure}[h]
\begin{minipage}{0.5\textwidth}
  \centering
  \includegraphics[width=5cm]{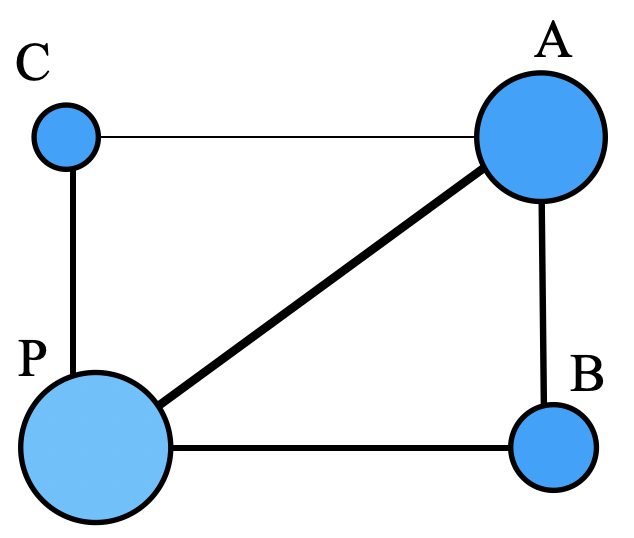}
  \figcaption{Diagram of simulation study 1}
  \label{figure:Simulation_1}
\end{minipage}
\begin{minipage}{0.5\textwidth}
\begin{center}
\def\@captype{table}
    \tblcaption{Setting of simulation study 1}
  \label{table:Simulation_1}
    \begin{tabular}{|l||c|c|} \hline
      Design & \# of Studies & \# of Studies \\ \hline
      P vs A & 3 & 9\\
      P vs B & 2 & 6\\
      P vs C & 2 & 6\\
      A vs B & 2 & 6\\
      A vs C & 1 & 3\\ \hline
      Total & 10 & 30 \\ \hline
    \end{tabular}
\end{center}
\end{minipage}
\end{figure}

There are four treatments (A-drug, B-drug, C-drug, and placebo [P]); the network diagram and the number of studies are shown in Figure \ref{figure:Simulation_1} and Table \ref{figure:Simulation_1}. We are interested in a comparison between the outcome of P and each of the three drugs. 

The first step is to generate the number of subjects in the studies ($n=n_A=n_B=n_C=n_P$) from the uniform distribution $\Uc(30,300)$ and the probability $p_P$ of occurrence of events in the placebo group from the uniform distribution $\Uc(0.05,0.65)$. We set $\bmu=(\mu_1,\mu_2,\mu_3)=(\mu_{A-P},\mu_{B-P},\mu_{C-P})=(0.398,0.702,0.866)$ and $\th=\th_1=\ldots=\th_6=0.3$. All off-diagonal elements of $\V(\th)$ are $\th/2$\cite{Higgins1996}. The next step is to generate the treatment effect within the study from $\Nc(\bmu,\V(\th))$. The probabilities of groups A, B, and C group can be derived from $p_P$ and the treatment effect. The final step is to generate the number of events from the binomial distribution $\Bc(p_i,n)$, where $i \in \{P, A, B, C\}$. The null hypotheses are $H_0:\mu_1=0.398$ for $A-P$, $H_0:\mu_2=0.702$ for $B-P$, and $H_0:\mu_3=0.866$ for $C-P$. The type 1 error is 5\%, i.e., we will confirm to be held the confidence interval to 95\%. We count the number of the confidence intervals that include the true value. The number of simulations is 10,000 replications. When we use the bootstrap method, it is computed on 1,000 replications and 1001 bootstrap samples\cite{Davidson2000}, as the numerical calculations take a great deal of time.

\begin{table}[h!]
  \begin{center}
\caption{Result of a simulation study (three comparisons)}
  \label{table:Simulation_1_res}
  \scalebox{0.8}[0.8]{
    \begin{tabular}{|c|c||cccc|cccc|cccc|} \hline
      \multicolumn{14}{|l|}{N=10}\\ \hline
      estimator & & W & W$^{BC}$ & W$_{boot}$ & W$^{BC}_{boot}$ & LR & LR$^{BC}$ & LR$_{boot}$ & LR$^{BC}_{boot}$ & S & S$^{BC}$ & S$_{boot}$ & S$^{BC}_{boot}$\\ \hline
      MLE & $\mu_1$ & 86.1 & 93.6 &  93.1 & 92.9 & 88.9 & 94.4 & 94.0 & 96.8 & 93.3 & 95.5 & 94.8 & 95.1\\
      MLE & $\mu_2$ & 86.8 & 93.8 & 92.9 & 92.9 & 89.2 & 94.8 & 93.5 & 97.6 & 93.7 & 96.1 & 95.1 & 95.1\\
      MLE & $\mu_3$ & 86.7 & 93.9 & 91.3 & 91.1 & 89.5 & 94.8 & 92.0 & 96.8 & 93.5 & 95.9 & 92.9 & 93.1\\ \hline
      REMLE & $\mu_1$ & 91.3 & 94.6 & 96.2 & 96.2 & 94.8 & 95.5 & 95.8 & 95.8 & 98.2 & 96.3 & 95.8 & 95.8\\
      REMLE & $\mu_2$ & 91.7 & 94.6 & 95.9 & 96.1 & 95.0 & 95.9 & 96.1 & 96.1 & 98.5 & 96.7 & 95.7 & 95.5\\
      REMLE & $\mu_3$ & 91.9 & 94.6 & 94.8 & 94.8 & 95.1 & 95.9 & 95.2 & 95.0 & 98.5 & 96.7 & 94.8 & 95.0\\ \hline
      \multicolumn{14}{|l|}{N=30}\\ \hline
      estimator & & W & W$^{BC}$ & W$_{boot}$ & W$^{BC}_{boot}$ & LR & LR$^{BC}$ & LR$_{boot}$ & LR$^{BC}_{boot}$ & S & S$^{BC}$ & S$_{boot}$ & S$^{BC}_{boot}$\\ \hline
      MLE & $\mu_1$ & 92.3 & 94.6 &  95.9 & 95.9 & 93.2 & 94.8 & 95.7 & 96.8 & 94.3 & 94.9 & 95.5 & 94.5\\
      MLE & $\mu_2$ & 92.3 & 94.6 & 94.6 & 94.7 & 93.1 & 94.8 & 94.6 & 96.3 & 94.4 & 95.0 & 94.5 & 94.5\\
      MLE & $\mu_3$ & 92.6 & 95.0 & 95.9 & 95.8 & 93.6 & 95.1 & 95.9 & 96.8 & 94.7 & 95.2 & 96.1 & 96.1\\ \hline
      REMLE & $\mu_1$ & 93.8 & 94.9 & 95.7 & 95.7 & 94.7 & 94.9 & 95.4 & 95.3 & 95.7 & 95.0 & 95.3 & 95.3\\
      REMLE & $\mu_2$ & 93.9 & 94.9 & 94.5 & 95.5 & 94.8 & 95.0 & 94.5 & 94.5 & 95.6 & 95.1 & 94.5 & 94.5\\
      REMLE & $\mu_3$ & 94.2 & 95.1 & 96.0 & 95.9 & 95.0 & 95.3 & 95.8 & 95.8 & 96.1 & 95.3 & 96.0 & 96.0\\ \hline
    \end{tabular}
    }
    \begin{tablenotes}\footnotesize
\item W: Wald; W$^{BC}$: Bartlett-type adjusted Wald; W$_{boot}$: Wald using bootstrap method; W$^{BC}_{boot}$: Bartlett-type adjusted Wald using bootstrap method; LR: Likelihood Ratio; LR$^{BC}$: Bartlett-type adjusted LR; LR$_{boot}$: LR using bootstrap method; LR$^{BC}_{boot}$: Bartlett-type adjusted LR bootstrap method; S: Score; S$^{BC}$: Bartlett-type adjusted Score; S$_{boot}$: Score using bootstrap method; S$^{BC}_{boot}$: Bartlett-type adjusted Score using bootstrap method
\end{tablenotes}
  \end{center}
\end{table}

The simulation results are in Table\ref{table:Simulation_1_res}. The all adjusted confidence intervals have been improved and closed to 95\%. The coverage probability of the Wald statistic is under 95\%, thus the confidence interval of the Wald statistic is narrower (details in Appendix A.12 Average confidence intervals in computer simulations). On the other hand, the confidence interval of the Score statistic is wider, since the coverage probability exceed 95\%. The confidence interval of the LR statistic is relatively stable because the bias terms are fewer than the Wald and Score statistics. In the case of MLE under $N=10$, the adjusted confidence interval based on the Wald statistic have been significantly improved. The confidence interval of the LR statistic using REML estimator seems less biased because the percent change in the adjusted statistic based on the unadjusted statistic in Appendix 12 in the supplemental material is very small as 3\%. As for the average simulation time of method, when $N=10$, our method took $42.3$ minutes (0.3 seconds per one simulation), whereas the bootstrap method took over about $4,000$ minutes (4 minutes per one simulation). Moreover, when under $N=30$, our method took $49.8$ minutes (0.3 seconds per one simulation), whereas the bootstrap method took over $5,000$ minutes (5 minutes per one simulation).

\subsection{Simulation 2}\label{subsec62}
\begin{figure}[h]
\begin{minipage}{0.5\textwidth}
  \centering
  \includegraphics[width=5cm]{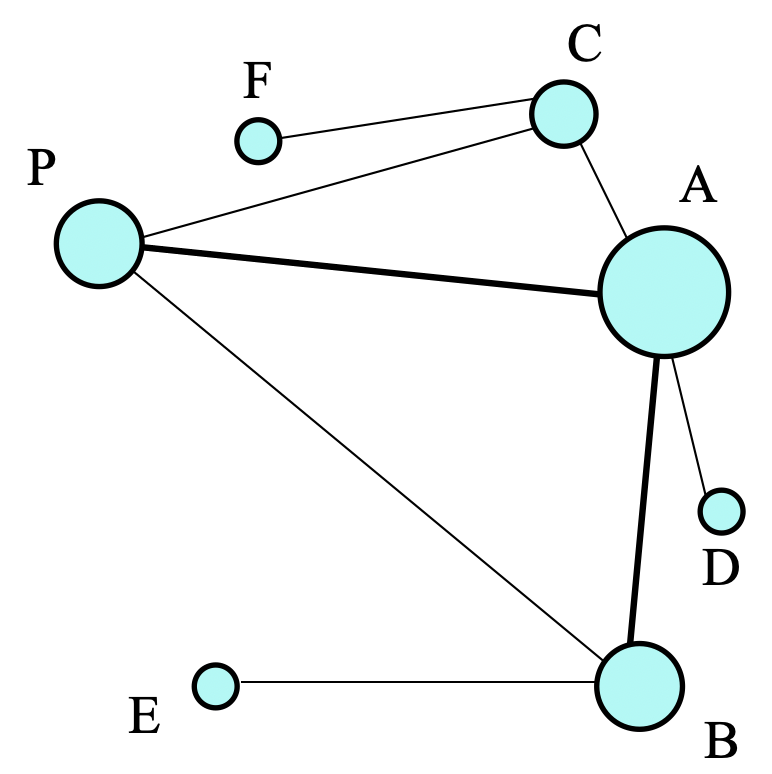}
  \figcaption{Diagram of simulation study 2}
  \label{figure:Simulation_2}
\end{minipage}
\begin{minipage}{0.5\textwidth}
\begin{center}
\def\@captype{table}
    \tblcaption{Setting of simulation study 2}
  \label{table:Simulation_2}
    \begin{tabular}{|l||c|c|} \hline
      Comparison & \# of Studies & \# of Studies \\ \hline
      P vs A & 6 & 4\\
      P vs B & 3 & 2\\
      P vs C & 3 & 2\\
      A vs B & 6 & 4\\
      A vs C & 3 & 2\\ 
      A vs D & 3 & 2\\ 
      B vs E & 3 & 2\\ 
      C vs F & 3 & 2\\  \hline
      Total & 30 & 20 \\ \hline
    \end{tabular}
\end{center}
\end{minipage}
\end{figure}

We assume that there seven treatments (A-drug, B-drug, C-drug, D-drug, E-drug, F-drug, and placebo [P]); the network diagram and the number of studies are shown in Figure 2 and Table 3. We are interested in a comparison between an outcome of P and all six drugs. We assume the almost same assumption as in Simulation 1, the probabilities of events are $p_A=0.6, p_B=0.65, p_C=0.7,  p_D=0.75,  p_E=0.8,  p_F=0.85, p_P=0.6$, and bootstrap samples are generated 501 times. The null hypotheses are $H_0:\mu_1=0$ for $A-P$, $H_0:\mu_2=0.093$ for $B-P$, $H_0:\mu_3=0.192$ for $C-P$, $H_0:\mu_4=0.301$ for $D-P$, $H_0:\mu_5=0.426$ for $E-P$, and $H_0:\mu_6=0.577$ for $F-P$. The network diagram and the number of studies are shown in Figure \ref{figure:Simulation_2} and Table \ref{figure:Simulation_2}. The total number of studies to $N = 20$ instead of $N = 10$.

\begin{table}[h!]
  \begin{center}
\caption{Covarage probability (\%) in a simulation study (six comparisons)}
  \label{table:Simulation_2_res}
  \scalebox{0.8}[0.8]{
    \begin{tabular}{|c|c||cccc|cccc|cccc|} \hline
      \multicolumn{14}{|l|}{N=20}\\ \hline
      estimator & & W & W$^{BC}$ & W$_{boot}$ & W$^{BC}_{boot}$ & LR & LR$^{BC}$ & LR$_{boot}$ & LR$^{BC}_{boot}$ & S & S$^{BC}$ & S$_{boot}$ & S$^{BC}_{boot}$\\ \hline
      MLE & $\mu_1$ & 87.3 & 93.3 & 95.2 & 95.5 & 88.8 & 93.8 & 95.2 & 98.3 & 91.4 & 94.6 & 94.5 & 94.5\\
      MLE & $\mu_2$ & 87.6 & 93.6 & 92.3 & 92.3  & 89.3 & 94.0 & 92.3 & 97.2 & 91.5 & 94.7 & 92.7 & 92.7\\
      MLE & $\mu_3$ & 87.4 & 93.4 & 94.1 & 93.4 & 89.2 & 94.2 & 93.8 & 95.8 & 91.7 & 94.9 & 93.8 & 93.8 \\
      MLE & $\mu_4$ & 87.4 & 93.4 & 93.1 & 93.1 & 89.0 & 94.1 & 93.4 & 96.9 & 91.5 & 95.0 & 93.8 & 93.8\\
      MLE & $\mu_5$ & 87.7 & 93.0 & 92.0 & 92.0 & 89.3 & 93.8 & 92.7 & 96.9 & 91.6 & 94.6 & 92.7 & 92.7\\
      MLE & $\mu_6$ & 87.4 & 93.1 & 95.5 & 95.5 & 89.0 & 93.7 & 96.5 & 98.6 & 91.4 & 94.6 & 97.9 & 97.6\\ \hline
      REMLE & $\mu_1$ & 93.2 & 94.7 & 95.8 & 95.5 & 95.0 & 95.3 & 95.8 & 95.8 & 96.9 & 95.9 & 95.8 &95.8 \\
      REMLE & $\mu_2$ & 93.4 & 94.9 & 93.7 & 93.7 & 95.1 & 95.5 & 94.1 & 94.1 & 96.8 & 95.9 & 94.4 &94.8 \\
      REMLE & $\mu_3$ & 93.6 & 95.1 & 95.5 & 95.5 & 95.4 & 95.7 & 95.5 & 95.5 & 97.2 & 96.2 & 94.4 &94.4 \\
      REMLE & $\mu_4$ & 93.6 & 95.1 & 93.8 & 93.8 & 95.4 & 95.7 & 94.8 & 94.8 & 97.0 & 96.3 & 94.8 &94.8 \\
      REMLE & $\mu_5$ & 93.2 & 94.6 & 94.4 & 94.4 & 95.0 & 95.4 & 94.8 & 94.8 & 96.7 & 95.9 & 95.5 &95.5 \\
      REMLE & $\mu_6$ & 93.2 & 94.6 & 97.6 & 97.6 & 94.9 & 95.4 & 97.6 & 97.6 & 96.9 & 95.9 & 97.9 &97.9 \\ \hline
      \multicolumn{14}{|l|}{N=30}\\ \hline
      estimator & & W & W$^{BC}$ & W$_{boot}$ & W$^{BC}_{boot}$ & LR & LR$^{BC}$ & LR$_{boot}$ & LR$^{BC}_{boot}$ & S & S$^{BC}$ & S$_{boot}$ & S$^{BC}_{boot}$\\ \hline
      MLE & $\mu_1$ & 90.2 & 93.9 &  94.8 & 94.8 &  91.3 & 94.3 & 95.1 & 96.7 &  92.7 & 94.8 & 95.3 & 95.1\\
      MLE & $\mu_2$ & 90.2 & 94.0 &  93.6 & 93.7 &  91.2 & 94.3 & 93.8 & 95.9 &  92.6 & 94.6 & 93.8 & 93.8\\
      MLE & $\mu_3$ & 90.0 & 94.0 &  93.7 & 93.7 &  91.1 & 94.3 & 93.7 & 96.3 &  92.6 & 94.7 & 94.2 & 94.1\\
      MLE & $\mu_4$ & 89.8 & 93.7 &  93.7 & 93.7 &  90.8 & 94.1 & 93.8 & 96.7 &  92.3 & 94.3 & 94.6 & 94.4\\
      MLE & $\mu_5$ & 90.3 & 94.0 &  93.1 & 93.1 &  91.4 & 94.4 & 93.5 & 95.9 &  92.7 & 94.8 & 93.2 & 93.1\\
      MLE & $\mu_6$ & 90.1 & 93.8 &  93.7 & 93.8 &  91.0 & 94.1 & 94.1 & 96.6 &  92.4 & 94.5 & 94.6 & 94.6\\ \hline
      REMLE & $\mu_1$ & 93.8 & 94.9 & 94.9 & 95.0 &  94.8 & 95.1 & 95.1 & 95.1 &  96.1 & 95.3 &95.3 & 95.3\\
      REMLE & $\mu_2$ & 93.6 & 94.8 & 93.8 & 93.7 &  94.8 & 95.0 & 93.8 & 93.8 &  95.9 & 95.2 & 93.8 & 93.8\\
      REMLE & $\mu_3$ & 93.7 & 94.8 & 94.2 & 94.1 &  94.8 & 95.1 & 94.5 & 94.5 &  96.1 & 95.2 & 94.2 & 94.4\\
      REMLE & $\mu_4$ & 93.5 & 94.5 & 94.9 & 94.9 &  94.5 & 94.9 & 95.3 & 95.0 &  95.9 & 95.2 & 95.0 & 95.0\\
      REMLE & $\mu_5$ & 93.8 & 94.8 & 93.9 & 93.8 &  94.8 & 95.2 & 93.7 & 93.7 &  96.1 & 95.4 & 93.9 & 93.9\\
      REMLE & $\mu_6$ & 93.5 & 94.7 & 94.6 & 94.6 &  94.7 & 95.0 & 94.2 & 94.2 &  96.0 & 95.3 & 94.2 & 94.2\\ \hline
    \end{tabular}
    }
  \end{center}
\end{table}

The simulation results are shown in Table \ref{table:Simulation_2_res}. Since the number of parameters has increased, the coverage probabilities of the confidence intervals have more bias. However, the adjusted confidence interval can be close to 95\%.

As for the average simulation time of each method, when $N=20$, our method took $230.1$ minutes (1.38 seconds per one simulation), whereas the bootstrap method took over about $23,000$ minutes (23 minutes per one simulation). Moreover, when $N=30$, our method took $283.7$ minutes (1.70 seconds per one simulation), whereas the bootstrap method took over $28,000$ minutes (28 minutes per one simulation).

We overview the entire results of simulation. Since the LR statistic is less biased compared to the Wald and score statistics and the adjusted LR statistic is the most stable, we recommend to use the adjusted LR statistic. The REMLE is more stable, We think it is better to use the REMLE than MLE.

\section{Applications to actual network meta-analysis}\label{sec7}
\subsection{Dual Antiplatelet therapy (DAPT)}\label{subsec71}
Palmerini et al.\cite{Palmerini2015} examined the mortality rate of three different extended durations of the dual Antiplatelet therapy (DAPT) for the definite or probable stent thrombosis rate via network meta-analysis of eight studies. The network diagram is shown in Figure 2. The study showed that short duration of DAPT decreases the mortality rate. We reanalyze the rate of the definite or probable stent thrombosis using the network meta-analysis. Let the reference group be 6 months, the variance-covariance matrix of random effect be the compound symmetry, and the sets of bootstraps samples be 10,001. The null hypotheses are set $H_0:\mu_1=0$ for 6 months vs 1 year, $H_0:\mu_2=0$ for 6 months vs 1 year over, and $H_0:\mu_3=0$ for 1 year vs 1 year over, because we are interested whether there are differences.

We show the results of re-analysis in Figure \ref{table:Palmerini2015result}. From the simulation results, the naive confidence intervals are too narrow or too wide due to bias effects. We confirmed the adjusted confidence intervals of the Wald statistic are wider. The adjusted confidence intervals of the LR statistic are also wider. The adjusted confidence intervals of the score statistic are narrower. 

\begin{figure}
  \label{table:Palmerini2015dig}
  \centering
  \includegraphics[width=5cm]{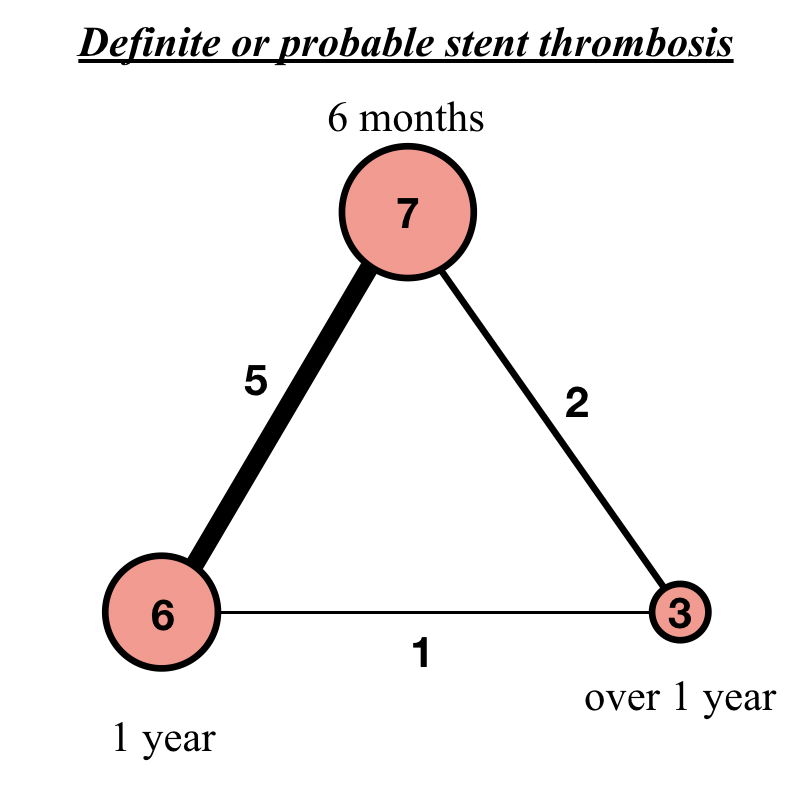}
  \caption{Palmerini2015 DAPT}
\end{figure}

\begin{figure}
  \centering
  \includegraphics[height=14cm,width=13cm]{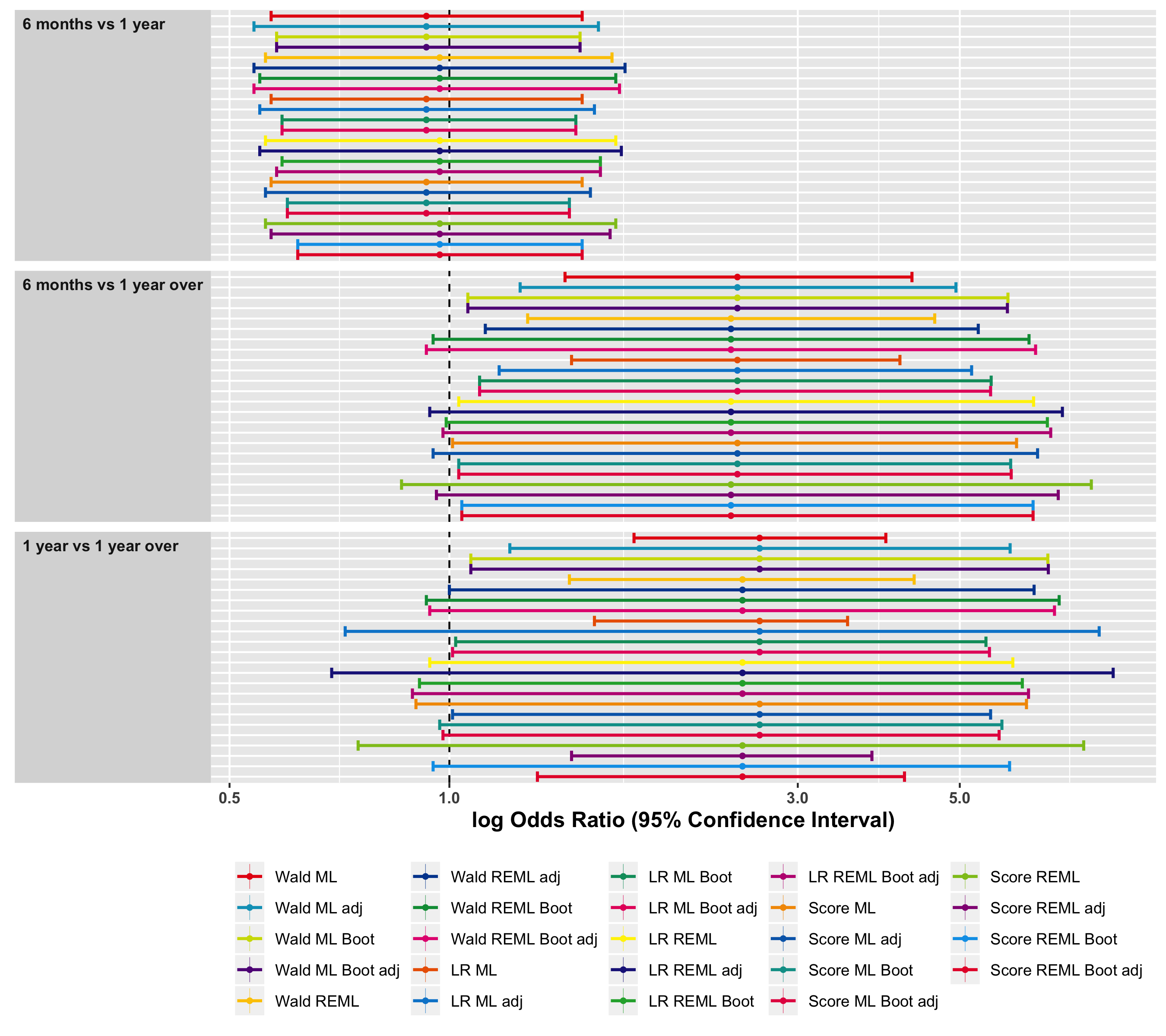}
  \caption{Palmerini2015 DAPT, definite or probable stent thrombosis}
  \label{table:Palmerini2015result}
\end{figure}
  
\section{Discussion}\label{sec8}
The problem with NMA using a small number of studies is that the confidence intervals cannot hold a nominal confidence level. To resolve this problem, we applied the Bartlett-type adjustment. Many Bartlett-type adjustment methods are based on the MLE. However, the NMA often uses the REMLE, which has not been extensively discussed in this context. We derived the novel Bartlett-type adjustment terms (\ref{eqn:Wcor}) - (\ref{eqn:Scor}) using not only MLE but also REMLE. 

Noma et al.\cite{Noma2018} proposed the Bartlett-type adjustment using $m$ bootstrap samples for NMA with a small number of studies. However, there is a problem when using bootstrap samples: the calculation load is very high, as the MLE or REMLE is derived by numerically solving a likelihood equation that is not closed-form. In addition, since the LR statistic and score statistic on the REMLE are not closed-form, and those confidence intervals also need to be derived numerically. When using the bootstrap method, we do not know which bias terms cause the bias. We proposed the closed-form confidence intervals of LR and score statistics and clarified the bias terms by deriving the Bartlett adjustment term analytically. In addition, we proposed a higher-order adjustment via the parametric bootstrap method. 

In the simulation results, because the LR statistic is less biased than the Wald and score statistics, and the adjusted LR statistic is the most stable, we recommend using the adjusted LR statistic. Because the REMLE is more stable, we think that it is preferable to use the REMLE rather than the MLE. In particular, the LR test statistic with the REMLE performed very well because it had less bias than the other statistics. We recommend using the LR test statistic with the REMLE, but just in case, we recommend applying the Bartlett-type adjustment to completely remove the bias. From demonstrations of actual studies, we confirmed that the adjusted confidence intervals were improved compared to the naive confidence intervals. 

In the future, we believe that NMA will become more popular\cite{Li2016}. Consequently, the Bartlett-type adjusted method will be used more frequently for the NMA with a small number of in which the number of studies is small. 

\bigskip
{\bf Acknowledgments.}\ \ 
The author is grateful to the editor, the associate editor and the referee for their valuable comments and helpful suggestions. The author would like to thank Associate Professor Hisashi Noma for his encouragement and helpful suggestions. This study was supported by a Grants-in-Aid for Scientific Research from the Japan Society for the Promotion of Science (Grant numbers: JP19H04074).

\bigskip
\noindent
{\bf Disclosures.}\ \ 

\noindent
None

\nocite{*}
\bibliography{manuscript}%



\end{document}